\documentclass[a4paper,fleqn,sort&compress]{cas-dc}

\usepackage[numbers, sort&compress]{natbib}
\usepackage[colorlinks]{hyperref}
\usepackage{graphics} 
\usepackage{amsmath}
\usepackage{amssymb}  
\usepackage{amsfonts} 
\usepackage{graphicx,float,wrapfig}
\usepackage{multirow}
\usepackage{makecell}
\usepackage{caption, subcaption}
\usepackage{bm}
\usepackage{soul,color}
\usepackage[ruled, vlined]{algorithm2e}
\usepackage{url}
\usepackage{booktabs}
\usepackage{xcolor}
\usepackage{multirow}
\usepackage{nomencl}

\usepackage{float}

\definecolor{deepfuchsia}{rgb}{0.76, 0.33, 0.76}
\definecolor{darkcoral}{rgb}{0.8, 0.36, 0.27}
\definecolor{cyan(process)}{rgb}{0.0, 0.72, 0.92}
\definecolor{etonblue}{rgb}{0.59, 0.78, 0.64}

\definecolor{glaucous}{rgb}{0.38, 0.51, 0.71}
\definecolor{green(munsell)}{rgb}{0.0, 0.66, 0.47}

\newcommand{\commentout}[1]{}

\def\tsc#1{\csdef{#1}{\textsc{\lowercase{#1}}\xspace}}
\tsc{WGM}
\tsc{QE}
\tsc{EP}
\tsc{PMS}
\tsc{BEC}
\tsc{DE}

\begin{document}
\let\WriteBookmarks\relax
\def\floatpagepagefraction{1}
\def\textpagefraction{.001}
\shorttitle{Flow-Aware Platoon Formation of Connected Automated Vehicles}
\shortauthors{Woo et~al.}

\title [mode = title]{Flow-Aware Platoon Formation of Connected Automated Vehicles in a Mixed Traffic with Human-driven Vehicles}                      



\author[1]{Soomin Woo}[orcid = 0000-0001-8326-3145]
\cormark[1] 
\ead{soomin.woo@berkeley.edu}  

\author[1]{Alexander Skabardonis}[] 
\ead{skabardonis@ce.berkeley.edu}  
\address[1]{416 McLaughlin Hall, University of California, Berkeley, California, USA 94720}

\cortext[cor1]{Corresponding author} 



\begin{abstract} 
Connected Automated Vehicles (CAVs) bring promise of increasing traffic capacity and energy efficiency by forming platoons with short headways on the road. However at low CAV penetration, the capacity gain will be small because the CAVs that randomly enter the road will be sparsely distributed, diminishing the probability of forming long platoons. Many researchers propose to solve this issue by \textit{platoon organization} strategies, where the CAVs search for other CAVs on the road and change lanes if necessary to form longer platoons. However, the current literature does not analyze a potential risk of platoon organization in disrupting the flow and reducing the capacity by inducing more lane changes. In this research, we use driving model of Cooperative Adaptive Cruise Control (CACC) vehicles and human-driven vehicles that are validated with field experiments and find that platoon organization can indeed drop the capacity with more lane changes. But when the traffic demand is well below capacity, platoon organization forms longer CAV platoons without reducing the flow. Based on this finding, we develop the \textit{Flow-Aware platoon organization strategy}, where the CAVs perform platoon organization conditionally on the local traffic state, i.e., a low flow and a high speed. We simulate the Flow-Aware platoon organization on a realistic freeway network and show that the CAVs successfully form longer platoons, while ensuring a maximal traffic flow.
%


\end{abstract}

\begin{graphicalabstract} 
\includegraphics[width=1\textwidth]{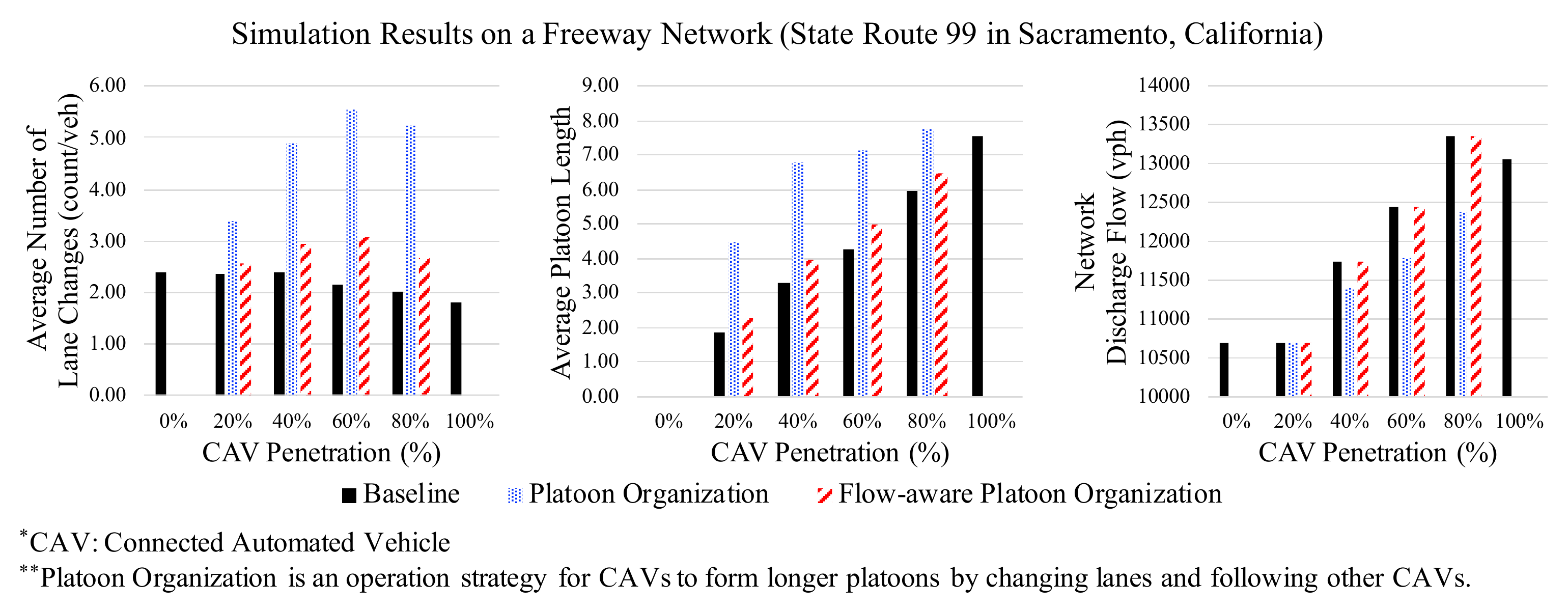}

\end{graphicalabstract}

\begin{highlights} 
    \item Connected Automated Vehicles (CAVs) promise to increase capacity by forming platoons.
    \item At low CAV penetration, the capacity gain will be small with few, short platoons.
    \item CAVs can change lanes to form platoons, but risk to disrupt flow and drop capacity.
    \item We propose a Flow-Aware strategy to form longer platoons with maximal traffic flow.
    \item We validate the Flow-Aware strategy by simulating traffic on a freeway network. 
\end{highlights}


\maketitle


\section{Introduction}
\label{sec:intro}
 
Cooperative Adaptive Cruise Control (CACC) is a vehicle technology that brings promise of greater road capacities and improved energy efficiency without investing on the road infrastructure, such as additional lanes or ramp metering controllers. Vehicles equipped with this technology are henceforth termed Connected Automated Vehicles, or CAVs for short. These vehicles monitor their speeds and gaps relative to their lead vehicles, and automatically adjust their motions in response. Moreover, CAVs can communicate with others of their kind nearby in real time and at high frequencies. With this communication, vehicle accelerations, lane-change maneuvers and other driving decisions can be shared across vehicles without the perception errors and reaction times associated with human drivers. These capabilities enable smaller vehicle headways than were previously possible (and thus larger road capacities) \cite{Shladover2012} and smaller air drag, improving energy efficiency \cite{Altinisik2015, Kaluva2020, Liu2020}. Note that the present research focuses on freeway setting to study the traffic impact of CAVs forming platoons without interruptions, such as signalized intersections.


\begin{figure}[h]
    \centering
    \includegraphics[width=.9\columnwidth]{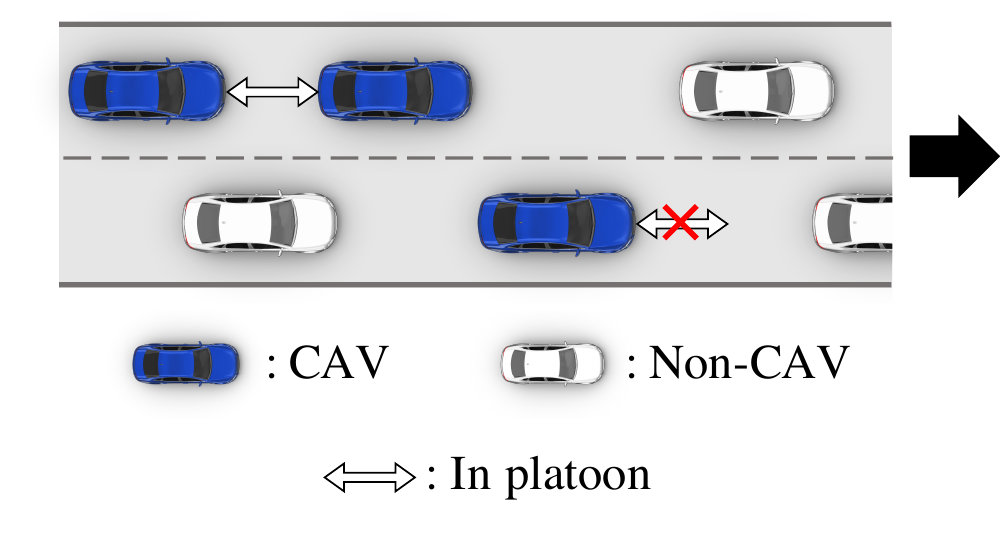}
    \caption{Challenges of CAVs at Low Penetration}
    \label{fig:illustration}
\end{figure}
 

Capacity gains are likely to be small, however, when the market penetration of CAVs is low in the early stage of implementation \cite{vanArem2006, Shladover2012, Liu2018_Modeling}. The sparse distribution of CAVs that randomly enter the road diminishes the probability of a CAV immediately following another of its kind, as shown in Figure~\ref{fig:illustration}. This introduces challenges when trying to form long platoons of CAVs so as to diminish the average headway in a traffic stream. To overcome this challenge, the CAVs can maneuver to organize the relative positions to each other on the road. The CAVs may have non-CAVs between them in the same lane or the CAVs may be in different lanes. A CAV can maneuver to follow another by accelerating, decelerating, or changing lanes. Then the gap between the consecutive CAVs must be reduced to form a platoon. In this present research, a term \textit{platoon organization} is defined to describe the strategies for CAVs to maneuver on the road to form longer platoons. 
 
Platoon organization has two potential impacts to the traffic. On the one hand, it can increase the number of CAVs in platoon, reducing the headways within the platoons and increasing the traffic capacity. On the other hand, it can induce more lane changes on the road that may disrupt the roadway flow \cite{Laval2006, Chen2018, Laval2007, Liu2020}. \commentout{The lane changes can also deactivate the connected driving mode of CAVs, causing further disruption \cite{Xiao2018}.}In other words, it is possible to negate the gain in capacity from longer platoons by the flow disruption from the induced lane changes. If platoon organization worsens capacity, it will be unfair for the society to suffer a lower capacity, while the drivers in CAVs enjoy the platooning benefits, such as fuel efficiency and driver comfort \cite{Altinisik2015, Kaluva2020, Liu2020}. Unfortunately, no literature has analyzed the traffic capacity under platoon organization, considering the impact of lane changes.   %

Many researchers propose to promote CAVs to form longer platoons with dedicated lanes for CAVs, where only CAVs are allowed to travel in the designated lanes and increase the probability of forming a platoon. However, their results on capacity vary. Some argue that traffic capacity increases with platoon organization, but do not consider the realistic impact of lane changes \cite{Ghiasi2017, Hua2020}. Ghiasi et al \cite{Ghiasi2017} analytically calculate that the capacity increases with dedicated lanes for CAVs, using a Markov Chain model. However, they assume that the CAVs enter the road already formed in longer platoons on the dedicated lanes and do not describe the impact of lane changes in capacity analysis. Hua et al \cite{Hua2020} use Cellular Automata model to show that the traffic capacity increases for all levels of CAV penetration with dedicated lanes for CAVs, however the lane change model is simple and does not model its flow disruption.

Some do not model the disruptive impact of lane changes, but still show that the capacity can drop at low CAV penetration because the dedicated lanes for CAVs are under-saturated \cite{Talebpour2017, Chen2017, Ye2018_Impact, Ma2019}. Talebpour et al \cite{Talebpour2017} show that the throughput is smaller at low CAV penetration but higher at high CAV penetration, if the CAVs optionally use the dedicated lanes for CAVs. However, they use a game-theory based lane change model, which is to be calibrated in future research \cite{Talebpour2015}. Chen et al \cite{Chen2017} analytically calculate that the capacity decreases at low penetration with dedicated lanes for CAVs, but capacity increases if the dedicated lanes are optional for the CAVs. They use a formulation derived from a single-lane capacity, recommending a future study to consider the effect of lane changes in the formulation. Ye et al \cite{Ye2018_Impact} show that throughput decreases in free-flow but increases in congestion at low CAV penetration, though their model does not describe the realistic impact of lane changes, such as the stochastic reaction of vehicles upstream of a lane-changing vehicle. Ma et al \cite{Ma2019} show that the dedicated lanes for CAVs decrease capacity at low CAV penetration but increases it over 40\% CAV penetration. However, they use modified Cellular Automata with a simple assumption for the CAVs to change lanes in a 'cooperative and instantaneous' manner.

Some model the impact of lane changes realistically, however do not analyze the traffic capacity under platoon organization \cite{Zhong2019, Xiao2020}. Zhong et al \cite{Zhong2019} use a CACC model (E-IDM) and a lane change model (MOBIL) that describe the flow disruption of lane changes. Xiao et al \cite{Xiao2020} implement an extensive model of lane change behavior \cite{Xiao2017_Modeling, Xiao2017_realistic} and simulate a mixed traffic of CAVs and non-CAVs on a freeway network. However, the analysis on traffic capacity is missing. At low CAV penetration, both studies report poorer traffic performance with dedicated lanes for CAVs because those lanes are under-saturated. The dedicated lanes will be more saturated at higher CAV penetration, however both studies only use the same demand for all CAV penetration levels and do not analyze the traffic flowing at capacity at high CAV penetration. Note that Xiao et al \cite{Xiao2020} observed speed reduction in the dedicated lanes for CAVs due to the lane changes to enter and exit the dedicated lanes.


To the best of our knowledge, it is still unclear if and how platoon organization increases or decreases the traffic capacity in a mixed traffic of CAVs and non-CAVs. To answer this question, we must analyze the traffic capacity, considering the impact of lane changes and using a platoon organization strategy that can saturate all lanes. Finally, it is necessary to design an operation strategy for CAVs to form longer platoons, while ensuring a maximal traffic flow. 

In this paper, we contribute to the current literature by revealing that platoon organization can reduce capacity and create congestion, by inducing lane changes that disrupt the flow. We use a driving model for CACC vehicle that is empirically verified from the real world data and conduct a sensitivity analysis on platoon organization at various traffic demands. We learn that when the demand is below capacity, platoon organization forms longer CAV platoons but does not reduce the flow. Based on this finding, we propose the \textit{Flow-Aware strategy of platoon organization} that forms longer CAV platoons and ensures maximal traffic flow without a capacity drop. We validate the Flow-Aware strategy by simulating it on a realistic freeway network.

This paper is structured as follows. In Section \ref{sec:researchappraoch}, we describe the microscopic traffic model to evaluate the traffic performance that considers the flow disruption of lane changes. We also propose a sample strategy of platoon organization that fully saturates all lanes at capacity. In Section \ref{sec:prelim_study}, we present a preliminary study to validate that platoon organization can drop the traffic capacity with more lane changes. The main experiments include the sensitivity analysis on platoon organization at various flow levels and the validation of the performance of the Flow-Aware Platoon organization strategy. In Section \ref{sec:methodology}, we describe the sensitivity analysis method, propose the Flow-Aware platoon organization strategy, and explain how we validate the Flow-Aware strategy on a freeway network. In Section \ref{sec:results}, we present the results of the sensitivity analysis and show that the Flow-Aware strategy produces longer platoons without flow disruption.  In Section \ref{sec:discussion}, we highlight the findings and discuss the shortcomings of this paper. Supplementary materials are given in the appendix, Section \ref{sec:appendix}.


\section{Research Approach}
\label{sec:researchappraoch}
In this work, we will conduct experiments to evaluate the impact of CAVs under platoon organization on traffic capacity and develop an CAV operation strategy that forms longer platoons and ensures a maximal traffic flow. Although the CAVs are not readily available in real settings, their driving behavior has been modeled via small scale experiments \cite{Nowakowski2010, Milanes2014, Milans2015, Lu2017}. We will use the driving models to simulate the mixed traffic of CAVs and non-CAVs. The following describes four features of a traffic model that this research requires to emulate platoon organization in a mixed traffic with CAVs and non-CAVs. The traffic models from the current literature are classified based on the four features and we will select one model to be used in the present study. In addition, we develop a sample strategy of platoon organization that saturates all lanes at capacity at all CAV penetration rates, eliminating the factor on capacity drop from unsaturated lanes.

\subsection{Microscopic Traffic Model}
\label{sec:metho_trafficmodel}
We need four features in a traffic model to emulate platoon organization. First, the model needs to describe mixed traffic composed of CAVs and non-CAVs at various penetrations of CAVs. Second, the model needs to be microscopic so as to depict the detailed operation of platoon organization, such as the number of CAVs in platoon. Third, the model needs to describe the impact of lane changes on roadway flow. Fourth, the model needs to be calibrated with experiments involving CAVs.

The current literature on traffic models for the mixed traffic is organized according to the four features in Table~\ref{tab:lit_on_micro_model}. Among them, the model in \cite{Liu2018_Modeling, Xiao2017_realistic} for mixed traffic satisfies the model requirements of the present study. Note that this model was calibrated for human and automated driving behaviors in \cite{Nowakowski2010, Milans2014} and the lane change model for human drivers was calibrated in \cite{Kan2019}. This present work uses the model in \cite{Liu2018_Modeling} as it is an improved model from \cite{Xiao2017_realistic}. Table~\ref{tab:micro_param} provides a list of parameters to be used in simulations from the model in \cite{Liu2018_Modeling}, used in the current study. In this paper, we simulate the traffic using Aimsun with external behavior enabled by MicroSDK.

Note that in this model, simulated drivers in CAVs change lanes manually because we assume the CAVs are equipped with CACC, which automates the car-following maneuvers but not the lane-change maneuvers. The model assumes that CAVs have a weaker motivation for discretionary lane changes than human drivers do because the CAVs prefer to stay in a platoon than to seek higher speeds. Note that when the brake is applied manually, the automatic car-following mode of a CAV is deactivated immediately and reverts to the manual driving mode. Additional description of the model is given in the appendix; see Section \ref{appendix:micro_model}.

\begin{table*}[cols=5,pos=h]
\caption{Literature on Microscopic Simulation of Mixed Traffic}
\label{tab:lit_on_micro_model}
\centering 
\begin{tabular*}{\textwidth}{l@{\extracolsep{\fill}}llll}
\toprule  
Publications & \makecell[c]{Mixed Traffic of\\CAVs and Non-CAVs} & \makecell[c]{Microscopic Model} & \makecell[c]{Lane Change Model} & \makecell[c]{Calibration with\\CAV Experiments} \\
\\[-1em]
\hline
\cite{Ghiasi2017, Ye2018_Modeling, Chen2017, Jin2018} & \checkmark & & & \\ 
\\[-1em]
\hline
\cite{Talebpour2016, Li2017} & \checkmark & \checkmark & & \\
\\[-1em]
\hline
\cite{Xiao2018} & \checkmark & \checkmark & \checkmark & \\
\\[-1em]
\hline
\cite{Liu2018_Modeling, Xiao2017_realistic} & \checkmark & \checkmark & \checkmark & \makecell[c]{\checkmark} \\
\bottomrule
\end{tabular*}
\end{table*}

\begin{table}[t]
\caption{Microscopic Driving Parameters}
\label{tab:micro_param}
\centering 
\begin{tabular}{l l l }
\toprule
Parameters & Non-CAV & CAV \\
\midrule
Average Reaction Time & 1.2s &0.4s \\ 
\midrule
Average Headway & 1.25s & 0.6 - 1.1s \cite{Nowakowski2010}\\
\midrule
Minimum Space Gap & 1.5m & 1.5m \\
\midrule
Average Maximum Acceleration &-4m/s$^2$ & -4m/s$^2$ \\
\midrule
Average Maximum Deceleration & 1.5m/s$^2$ & 1.5m/s$^2$ \\
\midrule
Vehicle Length & 4m & 4m \\
\midrule
\makecell[l]{Average of\\Maximum Desired Speed} & 110km/h & 110km/h \\
\midrule
Inter-platoon gap & NA & 1.5s\\
\midrule
\makecell[l]{Maximum Number of Vehicles\\in a Platoon} & NA & 10 \\
\bottomrule
\end{tabular}
\end{table}

\subsection{Platoon Organization Strategy} 
\label{sec:metho_samplestrategies}

\begin{figure}[h]
    \centering
    \includegraphics[width=1\columnwidth]{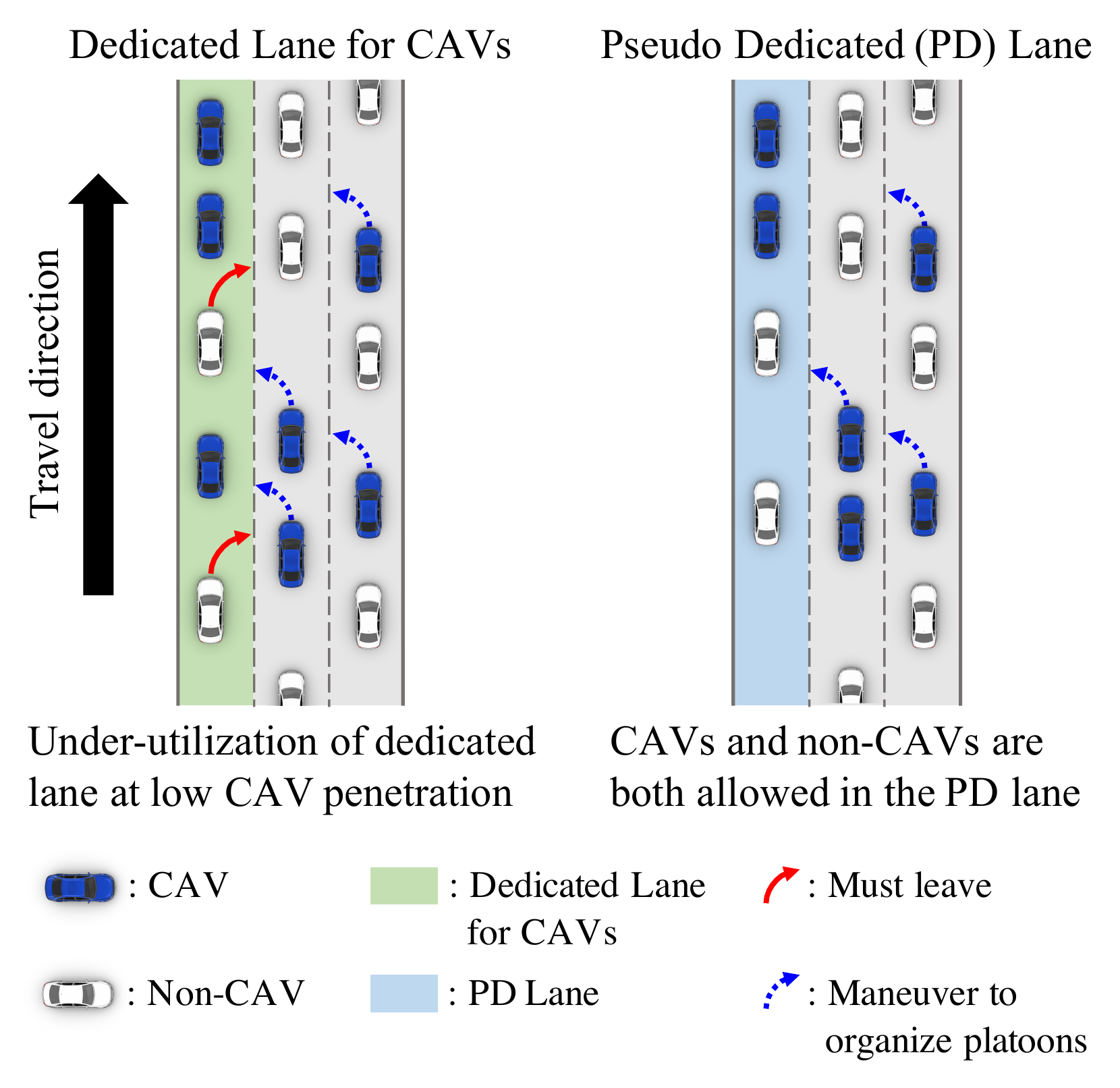}
    \caption{Pseudo Dedicated (PD) Lane}
    \label{fig:strategies}
\end{figure}
 

We develop a platoon organization strategy called pseudo dedicated (PD) lane, to avoid under-saturating a particular lane at high flow, which occurs with dedicated lanes for CAVs at low CAV penetration. Similar strategies have been proposed in the literature that allow both CAVs and non-CAVs on particular lanes \cite{Talebpour2017, Chen2017}. Figure~\ref{fig:strategies} explains the PD lane strategy in comparison to the dedicated lane for CAVs.  

The left figure shows the dedicated lane for CAVs, where only CAVs are allowed to travel on a dedicated lane.  At low penetration, the dedicated lane will not be saturated and worsen capacity.\footnote{The dedicated lanes for CAVs also pose an equity issue because a portion of public road is allocated only to people who can afford expensive CAVs \cite{Zhong2019}.} To isolate the effect of inefficient lanes in capacity estimation, the dedicated lane strategy is not used in this present study. In the pseudo dedicated (PD) lane strategy shown in the right figure, CAVs are motivated to maneuver into the PD lane for platoon organization, but non-CAVs are not banned from it. As the penetration increases, non-CAVs will naturally move away from the PD lane as more CAVs start to crowd it and travel slower than other lanes. We assume that due to the higher probability of forming long platoons, the CAVs have a stronger motivation to stay in the PD lane than the non-CAVs. This strategy thus guards against under-utilization of the lane at low market penetration of CAVs.

Throughout this study, we define \textit{baseline capacity}, as the traffic capacity of a given penetration rate of CAVs that do not perform platoon organization. We assume that a CAV can change a lane when it is the leader of a platoon. If a follower in a platoon were to change a lane, it must brake and split from its present platoon before the lane change. When a CAV immediately follows another CAV, it automatically starts to form a platoon. Note that we acknowledge a potential error in simulating the platoon organization with an assumption that the CAVs change lanes manually by the human drivers. The future work to improve this issue is described in Section \ref{sec:discussion} Discussion.

\section{Preliminary Study: Capacity Drop with Platoon Organization}  
\label{sec:prelim_study}
This section describes a preliminary study to test if platoon organization can drop capacity by inducing lane changes to disrupt the flow. We simulate the traffic on a homogeneous road segment, comparing the capacity with and without platoon organization. The results validate that when the CAVs change lanes to organize platoons, the traffic flow is reduced and the bottleneck capacity drops.  


\begin{figure}[h]
    \centering
    \includegraphics[width=.95\columnwidth]{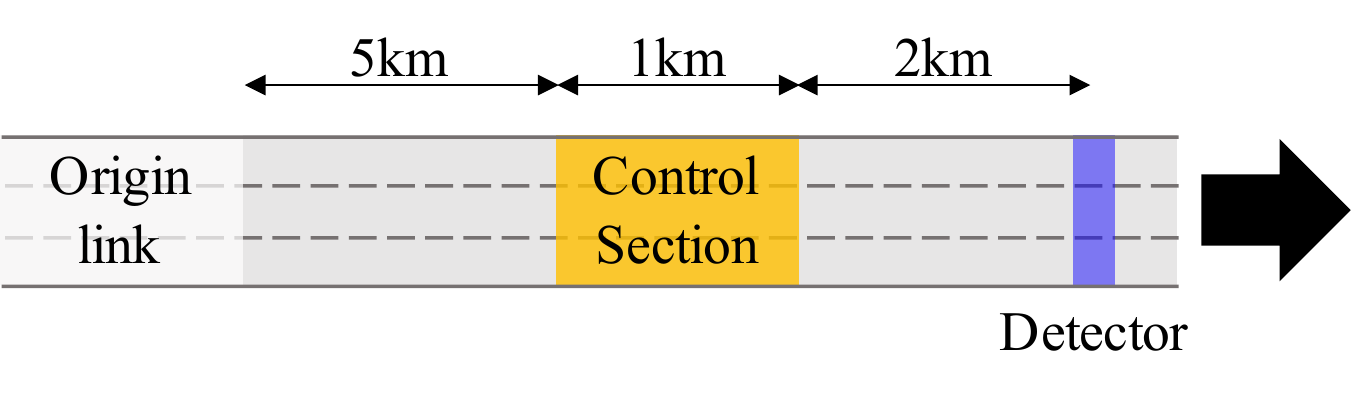}
    \caption{Homogeneous Road Segment}
    \label{fig:homogeneousroad}
\end{figure} 

This simulation experiment uses a 3-lane homogeneous road segment; refer to Figure~\ref{fig:homogeneousroad}. After the first stretch of 5km, CAVs execute the PD lane strategy in the control section of 1km. As the PD lane strategy allows both CAVs and non-CAVs, the PD lane is not regulated by infrastructural measures (like egress and ingress points) but by the lane targeting behavior of the CAVs. In the control section, the CAVs in short platoons change lanes to the PD lane to form longer platoons. CAV penetration levels are tested from 0 to 100\% at increments of 25\%. Vehicles are generated according to a Poisson process of a traffic demand rate and vehicle types are decided with Bernoulli trials with the probability of CAV type as the CAV penetration. 

The simulation procedure is as follows. First, the baseline capacity is estimated for each CAV penetration level without platoon organization. To estimate the baseline capacity, the traffic demand to the origin link as shown in Figure~\ref{fig:homogeneousroad} is increased until a queue forms in the link. The baseline capacity is measured as the maximum sustained flow discharged from the queue over a 60-minute period, at the flow detector shown in Figure~\ref{fig:homogeneousroad}. Second, the capacity of a bottleneck that forms in the control section under the platoon organization strategy is estimated for each penetration level of CAVs. For one hour, an input flow that produces the baseline capacity flow is used. The first 20 minutes is used as a warm-up period to create a stationary flow. In the next 40 minutes, the PD lane strategy is implemented in the control section and a queue forms in the control section. The bottleneck capacity is measured after the warm-up period for the 40-minute period as the maximum sustained flow discharged from the queue, measured at the flow detector.


\begin{figure}[h]
    \centering
    \includegraphics[width=0.99\columnwidth]{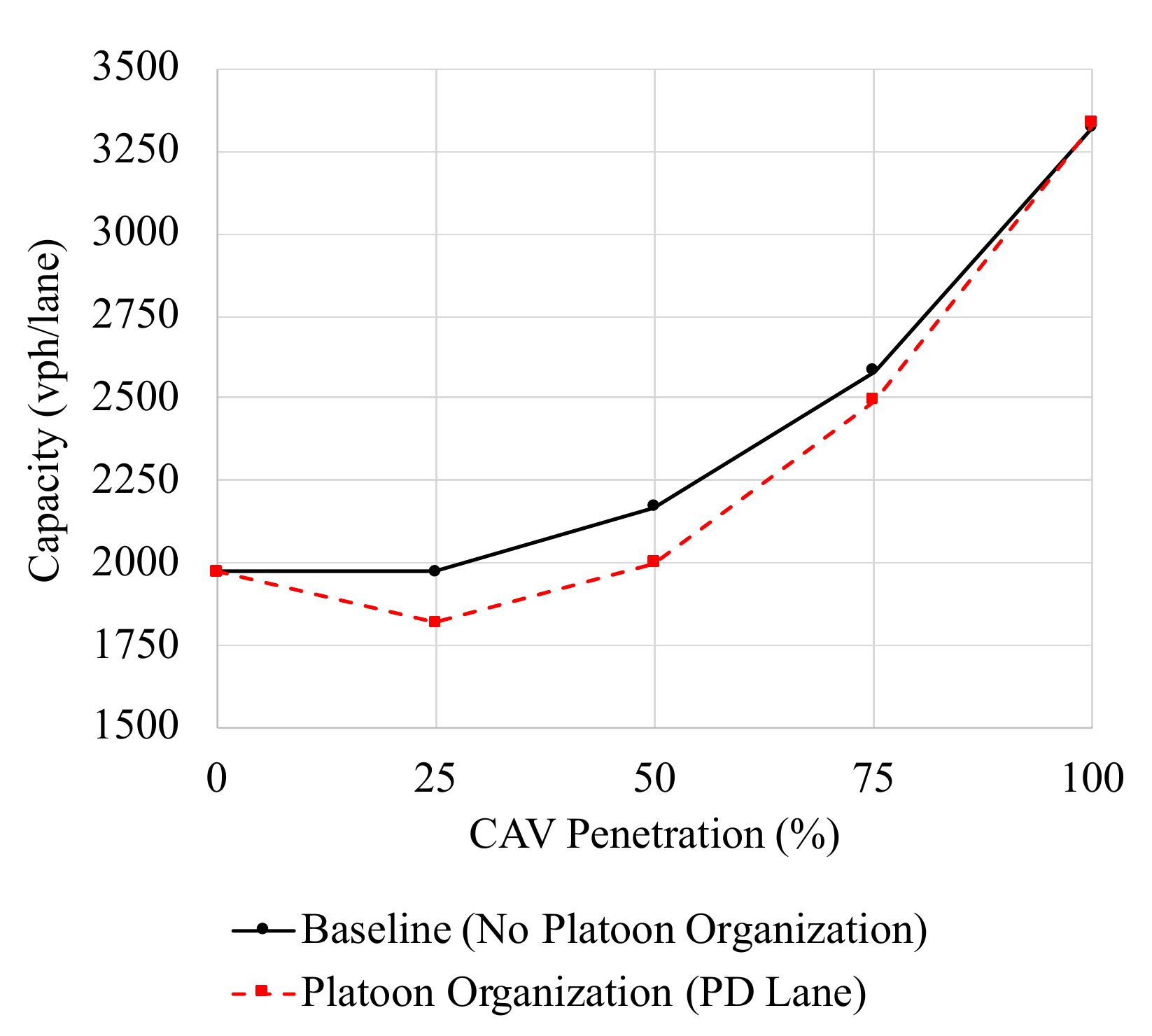}
    \caption{Capacity for Varying Market Penetrations of CAVs}
    \label{fig:capacitycomparison}
\end{figure} 


Figure~\ref{fig:capacitycomparison} shows the capacity with and without platoon organization as functions of CAV penetration. The baseline capacity without platoon organization is shown in a black solid line, which increases slowly at low CAV penetration up to 50\% and increases rapidly at higher penetration. We see that platoon organization worsens the capacity, as shown in a red dotted line. At 25\% CAV penetration, the PD lane strategy produces a capacity (around 1800 vph/lane) even less than the baseline capacity at 0\% penetration (around 2000 vph/lane). This means that capacity with CAVs under platoon organization can be worse than the capacity with no CAVs. At 50\% CAV penetration, the PD lane results in a capacity around 2000 vph/lane, similar to the baseline capacity at 0\% penetration. In other words, the PD lane can cancel the capacity improvement from having CAVs comprise half the vehicles on the road.
 
This negative effect of platoon organization on traffic capacity can be interpreted with the queuing theory, which describes that the output rate of a closed system is bounded by its input rate regardless of the dynamics in the system. We can define our closed system as the homogeneous road segment in Fig.~\ref{fig:homogeneousroad}, with one input and one output. Applying the theory to our experiment, it is impossible for the CAVs to increase the discharge flow (i.e., output rate) further than the input flow of baseline capacity (i.e., input rate), no matter how they maneuver on the road (i.e., platoon organization). Therefore, platoon organization cannot improve the capacity further than the baseline capacity. This must be true for any design of the closed system, whether it is a homogeneous road segment or a freeway network with multiple entrances and exits. Furthermore, the theory explains the capacity drop under platoon organization. When a system is at capacity, the output rate is bounded by the service rate. If the service rate decreases, the output rate decreases. Applying to our experiment, the platoon organization reduces the service rate by inducing more lane changes and inefficiently using the road. Therefore, the discharge flow (i.e., output rate) decreases. There is a need to prevent capacity drop under a naive implementation of platoon organization but still provide platooning benefits to the CAVs.  



\section{Methodology}
\label{sec:methodology} 
In this section, we describe the methodology of the two following experiments. We first conduct a sensitivity analysis on platoon organization with various levels of traffic demand, which provides the evidence in proposing the Flow-Aware strategy of platoon organization. The results provide the evidence in proposing the Flow-Aware strategy of platoon organization that ensures a maximal traffic flow and forms longer CAV platoons. Next, we validate the performance of the Flow-Aware strategy on a simulated freeway network.

\subsection{Sensitivity Analysis on Traffic Demand}
\label{sec:metho_second_experiment}  
The sensitivity analysis explores how platoon organization impacts the traffic performance under various levels of traffic demands. We use the homogeneous road segment in Fig.~\ref{fig:homogeneousroad}, where the CAV penetration is fixed at 50\% and input demands are tested at 1000, 1500, 2000, and 2500 vph/lane. The traffic is simulated for an hour. For the first 20 minutes, no platoon organization is implemented as a warm-up period. For the next 40 minutes, the PD lane strategy is be implemented in the control section. The simulation results are compared in terms of the average number of lane changes, the average platoon length, the platooning probability, and the discharge flow, which are all measured after the first 20 minutes. 
  
We define the \textit{platoon length}, $L_i(\tau)$, as the number of CAVs in a platoon that the $i$-th CAV is a member of at time $\tau$. The \textit{average platoon length}, $\Bar{L}$, is defined as the following:
\begin{align}
    \Bar{L} = \frac{\sum^{N_\text{CAV}}_{i=1} \max\limits_{\tau} \{L_i(\tau)\}}{N_\text{CAV}},
    \label{eq:average_platoon_length}
\end{align} 
where $N_\text{CAV}$ is the total number of CAVs that travel the network. The average platoon length computes the average of the maximum platoon lengths experienced by individual CAVs. Note that a CAV has $L_i(\tau) = 1$ if it is not connected to any other CAVs at time $\tau$. 

We also define the \textit{platooning probability}, $\mathsf{P}_\text{P}$, as the probability that a CAV is ever in a platoon with other CAVs as the following:
\begin{align}
    \mathsf{P}_\text{P} = \mathsf{P}(\max\limits_{\tau} \{L_i(\tau)\} > 1 ), \hspace{5mm} \forall i \in [1,{N_\text{CAV}}]
    \label{eq:platooning_probability}
\end{align}   

We calculate the average number of lane changes as the total number of lane changes that occur in the network, divided by the total count of vehicles including CAVs and non-CAVs. The discharge flow is measured at the flow detector.

\subsection{Flow-Aware Platoon Organization}
\label{sec:flow-aware-PO_propose}
We propose the \textit{Flow-Aware strategy of platoon organization} that forms longer platoons of CAVs but ensures maximal traffic flow. The main idea is for the CAVs to execute platoon organization conditionally on the traffic state. Since the traffic demand may be difficult to monitor in practice, we use speed and count to estimate the traffic condition and determine whether platoon organization is allowed. We do not enforce a specific location or time for the CAVs to change lanes for platoon organization, which is often how the ingress and egress points of special purpose lanes are operated. Instead, we determine if the traffic can absorb the disturbance from lane changes, based on the constant thresholds of flow and speed as $\rho_q$ (vph/lane) and $\rho_v$ (km/hr), respectively. 

Under the Flow-Aware strategy, a CAV gathers the average flow and speed measurements from the detectors nearby, as $\Bar{q}$ and $\Bar{v}$, respectively. The CAV evaluates a condition in Equation \eqref{flow-aware-condition} to check if the flow measure is lower than the threshold and the speed measure is higher than the threshold.
\begin{align}
    \{\Bar{q} < \rho_q\}  \land \{\Bar{v} > \rho_v\} \label{flow-aware-condition}
\end{align}
If the condition is satisfied, we assume that the traffic can `handle' the disturbances like lane changes and the CAV performs platoon organization, such as the PD lane strategy.

Note that when the threshold values, $\rho_q$ and $\rho_v$, are poorly calibrated, the Flow-Aware strategy may induce lane changes in a traffic with high demand and create unnecessary congestion. As the CAV penetration increases, the flow threshold, $\rho_q$, can be increased because the capacity improves with more CAVs and traffic can absorb more shock at a given demand level. Also the threshold values must be calibrated for different road networks as their geometries and capacities vary. For the freeway network studied in this paper, we calibrated the values of $\rho_q$ and $\rho_v$ for each CAV penetration rate by trial and error. We have reduced $\rho_q$ and increased $\rho_v$ so that an increase of lane changes does not decrease the network discharge flow significantly.

\subsection{Validation of Flow-Aware Platoon Organization on Freeway}
 \label{sec:metho_third_experiment}
\begin{figure}[h]
    \centering
    \includegraphics[width=.9\columnwidth]{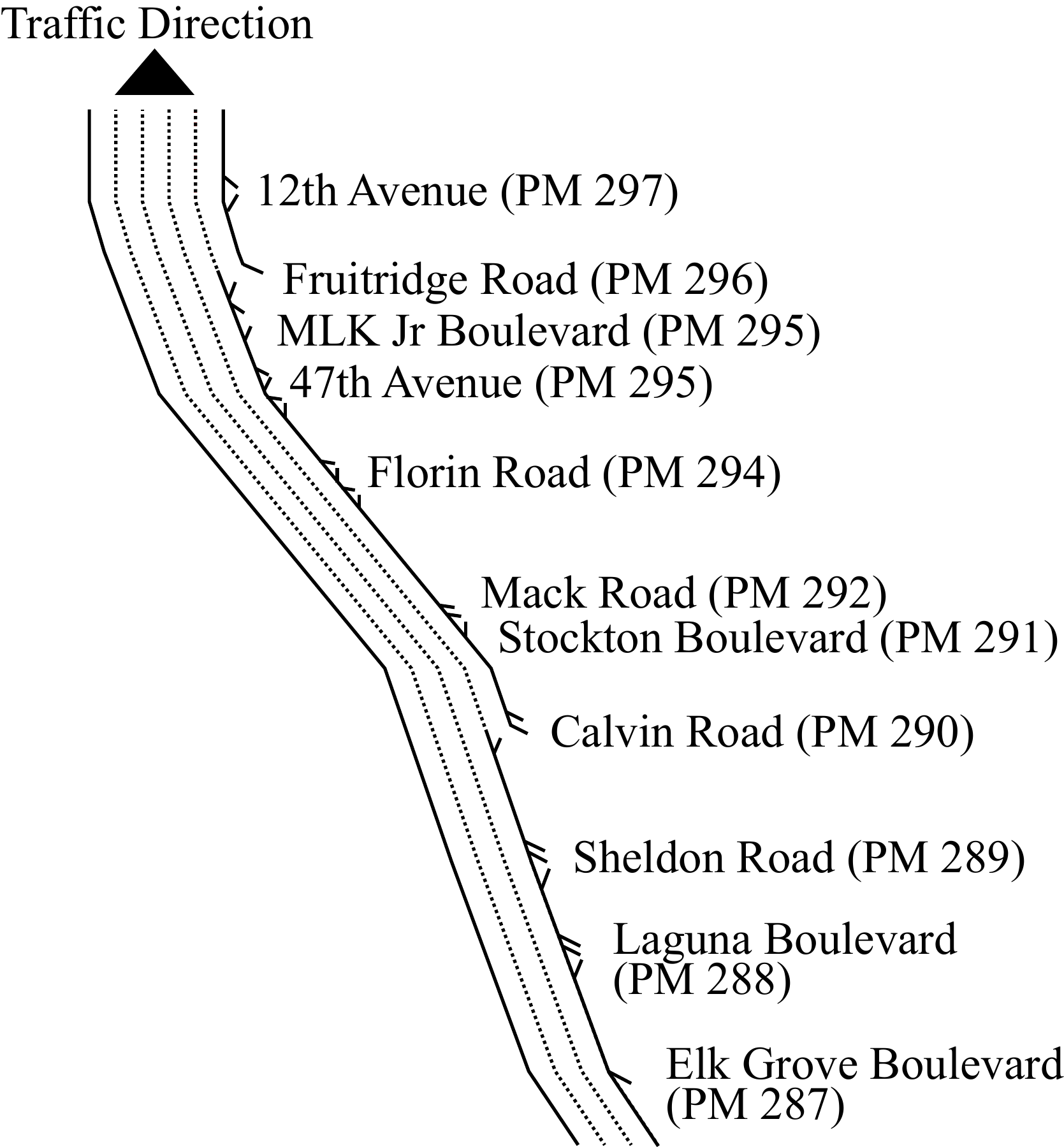}
    \caption{Simulated State Route 99 in Sacramento, California (PM: Post-mile)}
    \label{fig:sr99}
\end{figure} 

We test the Flow-Aware platoon organization strategy proposed in Section \ref{sec:flow-aware-PO_propose} by simulating it on a realistic freeway network and evaluating if the strategy ensures a maximal traffic flow and forms longer CAV platoons. We use a freeway network shown in Fig. \ref{fig:sr99}. This freeway network is the State Route 99 in Sacramento, California, covering the North-bound 10-mile (approximately 16-km) corridor. It starts at the Elk Grove Boulevard and ends at the 12th Avenue, including 16 on-ramps and 11 off-ramps. The network originally includes High-Occupancy Vehicle lanes, however we exclude them in this paper to isolate their effect on the capacity analysis. The traffic is simulated from 5AM to 9AM as the morning peak typically lasts from 6:30AM to 9AM. Bottlenecks recurrently form in the network, including one downstream of Elk Grove Boulevard due to high traffic demand. The free flow speed is around 105km/h for the freeway. 

We assume that on the freeway network, detectors are installed around 50 meters upstream of all on-ramps and off-ramps, as well as around 50 meters downstream of all on-ramps. The detectors aggregate data in 5-minute intervals and broadcast the latest information on the vehicle count and speed. The CAVs can gather information from detectors located within 1 to 2km range. This allows the CAVs under the Flow-Aware strategy to evaluate the local traffic condition with Equation \eqref{flow-aware-condition} and perform platoon organization accordingly. We also assume detectors to be installed at the end of each exit of the freeway, including the off-ramps and the main lanes after 12th Avenue, to collect the total discharge flow of the freeway network.

We test the CAV penetration from 0 to 100\% at increments of 20\%. At each entrance to the freeway, vehicles are generated according to the Poisson process of a traffic demand rate. At each exit of the freeway, vehicles depart the freeway as Bernoulli trials with a departure rate of the exit. We simulate three cases - the traffic without platoon organization as a baseline, the traffic with the PD lane strategy from Section \ref{sec:metho_samplestrategies} with no consideration to the traffic condition, and the traffic with Flow-Aware strategy from Section \ref{sec:results_second_experiment} with PD lane. 

There is no rigidly defined sections of ingress and egress to the PD lane because the PD lane allows both CAVs and non-CAVs. In the original PD lane strategy, the CAVs consider the leftmost lane of entire freeway as the PD lane. This strategy is an example of a naive platooning operation of the CAVs. In the Flow-Aware PD lane strategy, the traffic condition in Equation \eqref{flow-aware-condition} flexibly limits the execution of platooning maneuvers of CAVs. This strategy is an example of a platooning operation that controls the CAVs with awareness to the traffic flow.

The simulation procedure is as follows. First, we find the baseline demand for each CAV penetration level without platoon organization. We increase the network demand by scaling it with a constant factor, which we call \textit{demand scale}, until the speed contour shows the recurrent bottlenecks of the freeway network. Second, we use the baseline demand to simulate the network traffic under the PD lane strategy and the Flow-Aware PD lane strategy, at each CAV penetration rate. The results are compared in terms of the network discharge flow, average number of lane changes, and average platoon length. The network discharge flow is measured as the sum of discharge flows measured at the detectors on the end of each freeway exit. Note that we analyze the results for each CAV penetration rate (not between different penetration rates), because the baseline traffic is simulated at different demand scale for each penetration rate.

\section{Results} 
\label{sec:results}
In this section, we find that at a low demand the CAVs can form longer platoons without reducing the discharge flow under platoon organization. Therefore, the Flow-Aware strategy conditions the CAVs to perform platoon organization only under a low demand condition so that the output flow does not drop with the induced lane changes. We confirm the performance of the Flow-Aware strategy of platoon organization by simulating the traffic on a realistic freeway network. We show that the Flow-Aware strategy ensures a maximal traffic flow and enhance CAV platooning performance.

\subsection{Sensitivity Analysis on Traffic Demand}
\label{sec:results_second_experiment}

\begin{table*}[width=\textwidth,cols=10, pos=t]
\caption{Platoon Organization at Various Traffic Demands (Homogeneous Road in Fig. \ref{fig:homogeneousroad}, CAV Penetration of 50\%)}
\label{tab:various_input_flow}
\centering 
\begin{tabular*}{\textwidth}{l@{\extracolsep{\fill}}lllllllll}
\toprule   
\multirow{2}{*}{\makecell[l]{Traffic Demand\\(vph/lane)}} & \multicolumn{2}{l}{\makecell[l]{Average Number of \\Lane Changes\\(count/veh) $^{**}$}} & \multicolumn{2}{l}{\makecell[l]{Average\\Platoon Length, $\Bar{L}$,\\(CAV count)}} & \multicolumn{2}{l}{\makecell[l]{Platooning\\Probability, $\mathsf{P}_\text{P}$ $^{***}$}} & \multicolumn{3}{l}{\makecell[l]{Discharge Flow (vph/lane)}} \\
\\[-1em]
\cline{2-10}
\\[-1em]
&  B & PO & B & PO  & B & PO & \makecell[l]{$Q_B$} & \makecell[l]{$Q_{PO}$} & \makecell[l]{$\dfrac{Q_{QO} - Q_B}{Q_B}$} \\
\midrule
1000 & 0.05 & 0.49 & 2.6   & 5.0 & 0.73 &  0.89 & 1006.3 &1006.7 & {\color{blue}{0.04\%} } \\
\midrule
1500&0.07 & 0.51 & 3.0   & 5.1 & 0.74 &  0.89 & 1483.4 &1478.3 & {\color{blue}{-0.34\%}} \\
\midrule
2000 & 0.05 & 0.53 & 2.8   & 5.1 & 0.74 &  0.89 & 1961.7 & 1924.6 & -1.89\%  \\
\midrule
2500 & 0.04 & 0.56 & 3.3   & 5.4 & 0.78 &  0.91 & 2260.6 &2045.0 & {\color{red} -9.54\%} \\
\bottomrule 
\multicolumn{10}{l}{\makecell[l]{$^*$B: Baseline, PO: Platoon Organization, $Q_B$: Baseline discharge flow, $Q_{PO}$: Discharge flow under PO}} \\
\multicolumn{10}{l}{\makecell[l]{$^{**}$Note that the number of lane changes increase both for CAVs and non-CAVs.}} \\
\multicolumn{10}{l}{\makecell[l]{$^{***}$Note that the platooning probability $\mathsf{P}_\text{P}$ around 0.75 in the baseline makes sense for the baseline traffic of 50\% CAV\\penetration. Here is a simple explanation. Denote the CAV penetration rate as $P_\text{CAV}$. Then the platooning probability, $P_p$ can\\be expressed as the probability that given the subject vehicle is a CAV, its leader and/or follower vehicle is a CAV, i.e.,\\$P_p = 1- (1-P_\text{CAV})^2$. For $P_\text{CAV} = 0.5$, $P_p = 0.75$. The simulation well approximates this probability.}}
\end{tabular*}
\end{table*} 

We analyze the traffic performance at various traffic demands under the PD lane strategy on a homogeneous road segment at CAV penetration of 50\%. Table~\ref{tab:various_input_flow} presents the average numbers of lane changes, average platoon lengths, platooning probability, and the discharge flows at various traffic demands. Note that without platoon organization, a demand of 2,000 vph/lane results in a high but free flow. Without platoon organization, a demand at 2,500 vph/lane results in the capacity flow. We see that at all demand levels, platoon organization induces more lane changes (with higher average numbers of lane changes in the third column) and enhances platooning performance (with higher average platoon lengths in the fifth column and higher platooning probability in the seventh column). However, platoon organization affects the discharge flow very differently depending on the traffic demand; refer to the last column of Table \ref{tab:various_input_flow}. 

At low demands (1,000 to 1,500 vph/lane), more lane changes under platoon organization do not change the discharge flow significantly. At moderately high demand (2,000 vph/lane), the lane changes under platoon organization breaks down the traffic and reduces the discharge flow by 1.89\%. At a high demand (2,5000 vph/lane), the discharge flow reduces significantly by 9.5\%. Therefore, platoon organization can be used when there is low demand, where the disturbance from lane changes can be absorbed easily. At a moderately high demand (2,000 vph/lane), platoon organization can also break down the traffic and cause a bottleneck. At a high demand, platoon organization is ill-advised. The Flow-Aware platoon organization is designed with this mechanism by estimating the traffic demand based on speed and count measurements and elongating the platoon lengths without flow disruption.

\subsection{Flow-Aware Platoon Organization on a Freeway Network}
\label{sec:results_third_experiment}

\begin{figure*}
    \centering
    \includegraphics[width=1\textwidth]{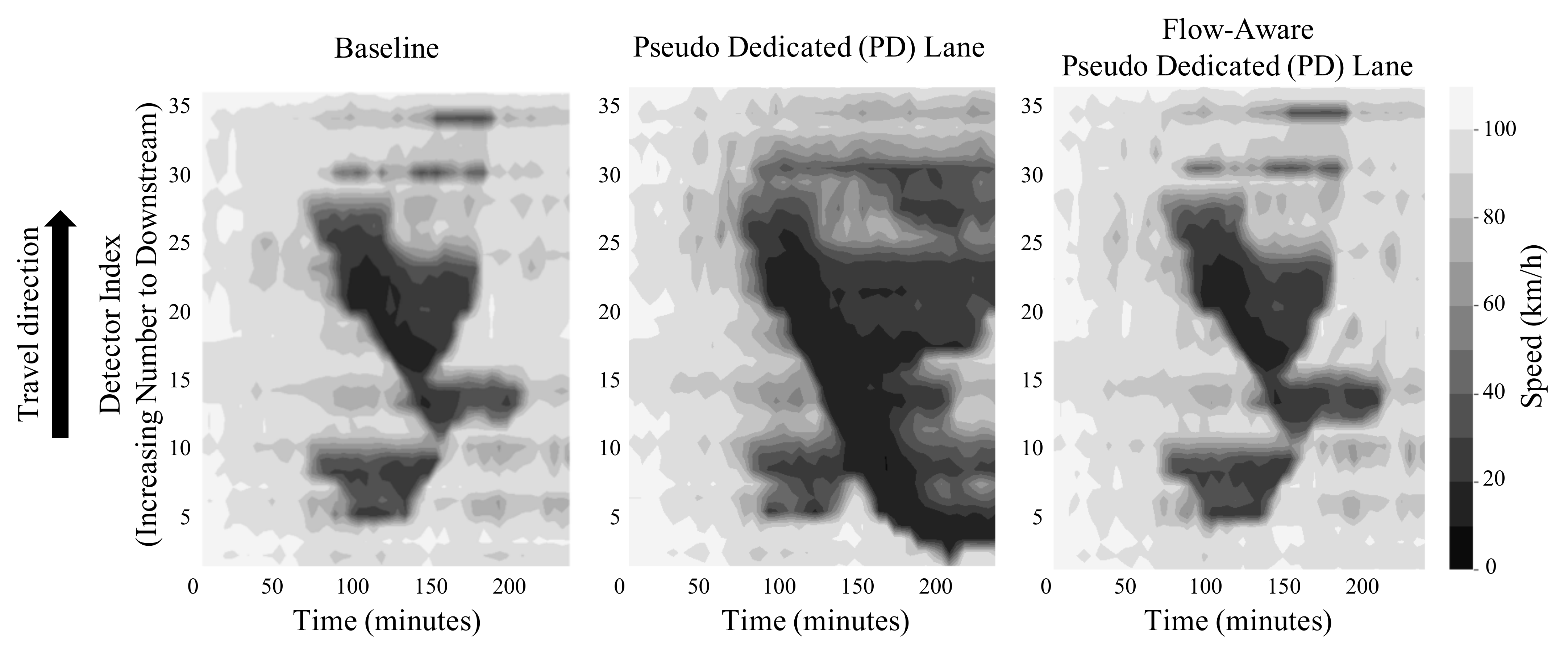}
    \caption{Speed Contours (Freeway Network in Fig.~\ref{fig:sr99}, CAV Penetration of 40\%): the PD lane strategy reduces the speed significantly from the baseline, whereas the Flow-Aware strategy with PD lane does not.}
    \label{fig:speedcontour_sr99}
\end{figure*} 

\begin{figure*}
    \centering
    \includegraphics[width=1\textwidth]{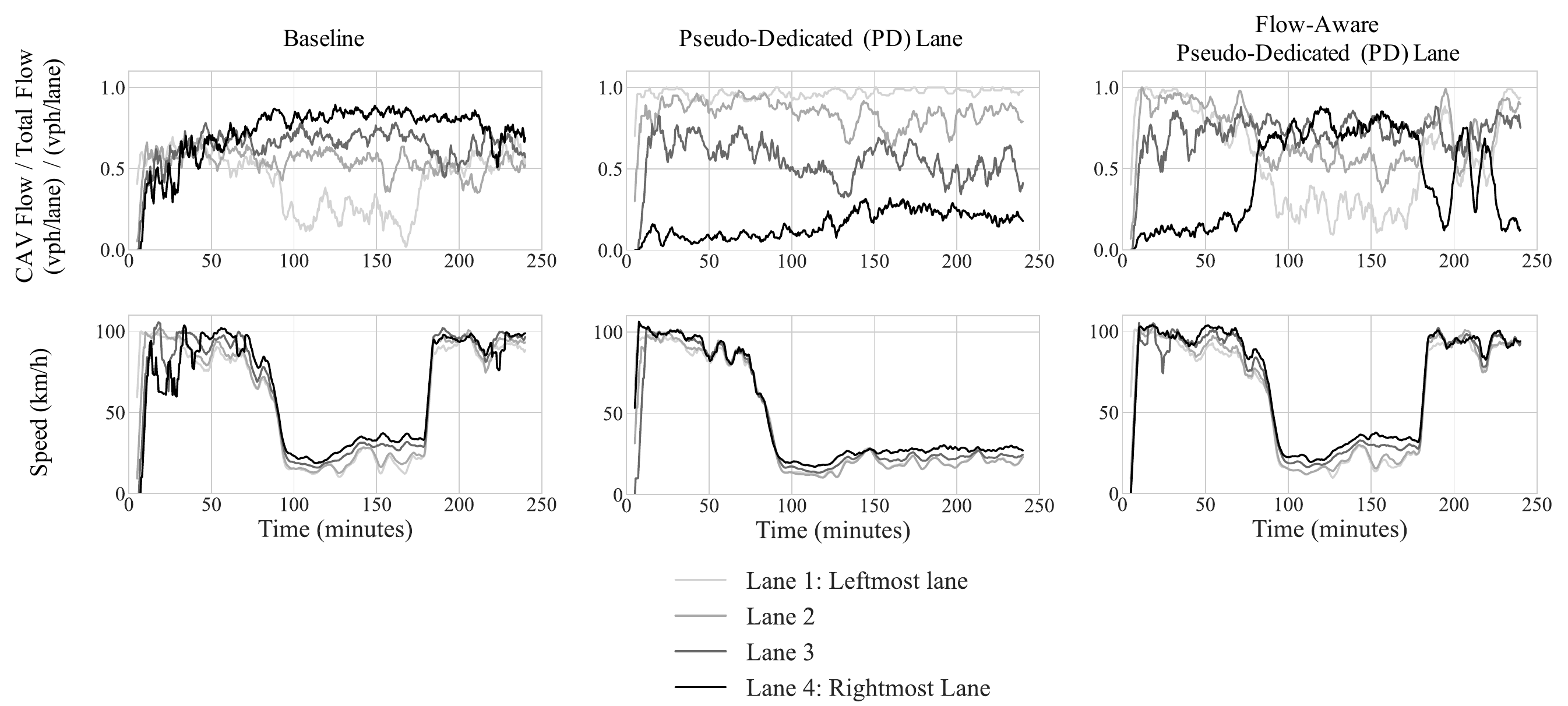}
    \caption{Ratio of CAV Flow to Total Flow (Freeway Network in Fig.~\ref{fig:sr99}, CAV Penetration of 40\%): the PD lane strategy distributes the CAVs to the leftmost PD lane regardless of the traffic condition, whereas the Flow-Aware strategy with PD lane only does so in a free flow.}
    \label{fig:lanebylane_cav_flowratio_sr99}
\end{figure*} 

We emulate the traffic on a freeway network to validate the strategy of Flow-Aware platoon organization. We assume that the CAVs perform PD lane strategy when the condition in Equation \eqref{flow-aware-condition} is met at any post-mile of the freeway. We first validate the traffic simulation with Fig.~\ref{fig:speedcontour_sr99}, which shows three speed contours for the baseline, the PD lane strategy, and the Flow-Aware PD lane strategy at 40\% CAV penetration. The simulation time is on the x-axis and space on the y-axis with detector index of an increasing number downstream. We observe that bottlenecks form in the baseline due to traffic demand larger than the network capacity. The PD lane strategy induces lane changes of CAVs regardless of the traffic condition and reduces speed significantly compared to the baseline. However, the Flow-Aware PD lane strategy results in a speed contour very similar to the baseline, indicating that the speed is not reduced significantly due to the platoon organization.

\begin{figure}[h]
    \centering
    \includegraphics[width=1\columnwidth]{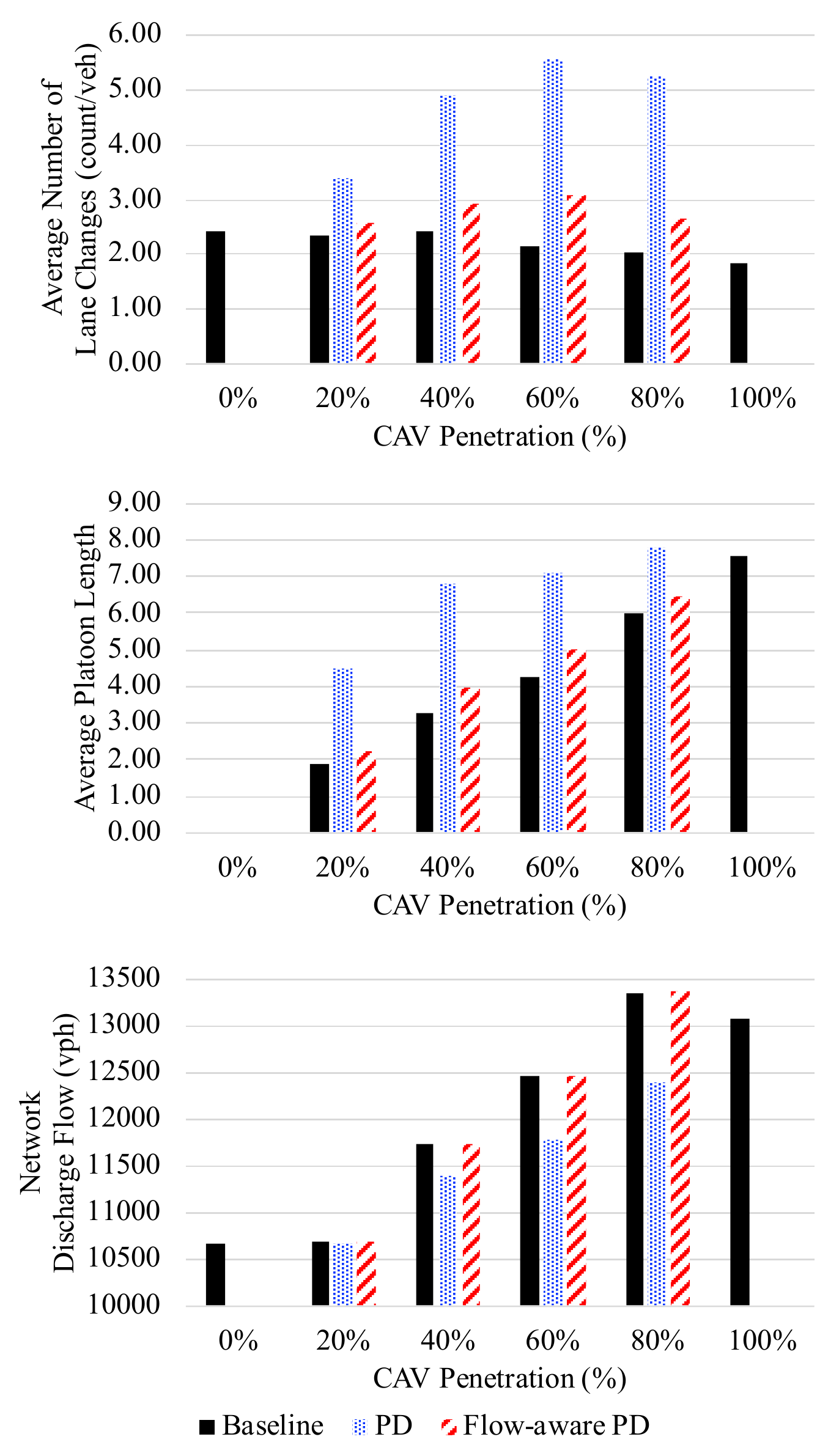}
    \caption{Increase of Lane Changes and Platoon Lengths without Flow Disruption under Flow-Aware Platoon Organization}
    \label{fig:totalcomparison}
\end{figure} 

We also observe the distribution of CAVs across the lanes under platoon organization with Fig.~\ref{fig:lanebylane_cav_flowratio_sr99} to validate the simulation further. The figure describes the traffic with 40\% CAVs in the section immediately upstream of the second on-ramp at Florin Road (shown in Fig.~\ref{fig:sr99}) or around the detector index 22 (shown in Fig.~\ref{fig:speedcontour_sr99}), where a bottleneck queue reduces the speed significantly. In Fig.~\ref{fig:lanebylane_cav_flowratio_sr99}, the x-axis is the simulation time, the y-axes for the top figures show the CAV flow ratio, i.e., the ratio of the flow of CAVs to the flow of all vehicle types, and the y-axes for the bottom figures show the speed per lane. The lanes are differentiated by the shade, where the lightest line corresponds to the leftmost (or PD) lane. The baseline, the PD lane strategy, and the Flow-Aware PD lane strategy are shown on the left, the center, and the right, respectively.
  
The baseline results on the left column show that when this section is in a bottleneck queue from around 80 minutes to 170 minutes, the CAVs are more concentrated on the right lanes than the left lanes. The CAVs make up about 80\% in the rightmost lane, whereas they make up about 20\% in the leftmost lane. This is possibly because the human driven vehicles have moved toward the leftmost lanes to avoid the conflict with the on-ramp merging flow.

The results under PD lane strategy on the center column show an opposite distribution of CAVs across lanes to the baseline results. The CAVs are highly concentrated in the leftmost PD lane all throughout the simulation. With the naive implementation of the PD lane strategy, the CAVs join the PD lane regardless of the traffic condition. This cause a longer duration of congestion.  

The results under the flow-aware PD on the right column are interesting, as the distribution of CAVs are similar to the PD lane during free flow, but similar to the baseline during congestion. During free flow (when speed is around 100km/h), the CAVs concentrate on the leftmost PD lane as the traffic condition in Equation \eqref{flow-aware-condition} is met and their maneuvers do not disrupt the flow. During congestion (from around 80 minutes to 170 minutes), the CAVs refrain from platoon organization because the condition in Equation \eqref{flow-aware-condition} is not satisfied. The CAVs concentrate higher in the right lanes like in the baseline case. Because of the restricted lane changes to the PD lane under congestion, the queue is dissipated at a similar time to the baseline.

We evaluate the traffic performance under the Flow-Aware platoon organization in comparison to the baseline and the PD lane strategy with Fig.~\ref{fig:totalcomparison}. The detailed values for the figure are found in Table.~\ref{tab:network_result_v3} in the Appendix Section \ref{appendix:flowaware-results}. Three figures present the average number of lane changes, average platoon length, and the network discharge flow, as functions of CAV penetration rates. At each penetration rate, the solid black bar shows the baseline result, the blue dotted bar shows the PD lane result with no condition on traffic, and the red dashed bar shows the result of Flow-Aware strategy with PD lane. Note that there is no result for the PD lane strategy and the Flow-Aware PD lane strategy for 0\% and 100\% CAV penetration rates because no CAV needs to be organized for longer platoons. The average platoon length is not applicable at 0\% CAV penetration. The demand scales are 100\%, 100\%, 110\%, 117\%, 125\%, and 127\% for CAV penetration rates of 0\%, 20\%, 40\%, 60\%, 80\%, and 100\%, respectively.

The top figure in Fig.~\ref{fig:totalcomparison} shows that more lane changes are induced under both PD lane strategy and the Flow-Aware PD lane strategy, compared to the baseline. The PD lane strategy naively induces more lane changes than the Flow-Aware strategy, without checking on the flow conditions. The middle figure in Fig.~\ref{fig:totalcomparison} shows the PD lane strategy increases the platoon length the most, but the Flow-Aware PD lane strategy also significantly increases the average platoon length (by 8 to 20\% as shown in Table.~\ref{tab:network_result_v3}).  

The bottom figure in Fig.~\ref{fig:totalcomparison} shows the network discharge flow. Compared to the baseline, the PD lane strategy significantly reduces the network flow (up to 7\% at 80\% as shown in Table.~\ref{tab:network_result_v3}). However, the Flow-Aware PD lane strategy does not decrease the network flow significantly (only up to 0.1\% as shown in Table.~\ref{tab:network_result_v3}). The Flow-Aware PD lane strategy successfully forms longer platoons while ensuring the maximum level of traffic flow. In other words, the CAVs can form longer platoons without trading off the level of traffic flow. Agreeing with the preliminary analysis in Section \ref{sec:prelim_study}, platoon organization does increase the network flow further than the baseline. 

Note that the network discharge flow in the baseline is lower at 100\% than at 80\%. Under the current simulation settings, the CAVs form and remain in a platoon if possible. The preference of CAV drivers to choose between forming a platoon and traveling in a faster lane is not well modeled or calibrated in this study; please refer to the future work in Section \ref{sec:discussion}) Discussion. At full CAV penetration, the CAVs form platoons soon after entering from the on-ramps because they find other CAVs immediately and stay in platoon on the right main lanes. So the highly dense traffic on the right lanes create conflicts with other merging flows downstream, resulting in a lower discharge flow than 80\% CAV penetration. This issue is out of scope for the current study, as we compare the baseline at each CAV penetration rate, not across different penetration rates. However, please refer to \cite{Liu2018_Impact} for further investigation on the need for a merging assistance system for a mixed traffic with CAVs and non-CAVs.

\section{Discussion}
\label{sec:discussion}
 
In this paper, we examine the potential problem of disrupting the traffic flow and causing unnecessary congestion with a naive platooning operation of CAVs. At low penetration, the CAVs will not deliver a large gain in the traffic capacity and the fuel economy because they are likely to form only few, short platoons. The CAVs can operate under a \textit{platoon organization} strategy, where they can maneuver to follow other CAVs on the road by changing lanes and form longer platoons. However, a poorly designed strategy of platoon organization can deteriorate the traffic capacity by inducing lane changes that disrupt the flow. We explore this issue by implementing the realistic driving models of CACC and human-driven vehicles, which are validated by field experiments and model the disruptive effects of lane changes. The key findings from this research are as below.
 
\begin{itemize}
    \item CAVs can form longer platoons by platoon organization, but may induce more lane changes on the road. At low traffic demand, the induced lane changes do not affect the traffic performance significantly as the traffic flows below capacity. However at high traffic demand, the induced lane changes can drop the capacity and create unnecessary traffic congestion.
    \item We propose the \textit{Flow-Aware strategy of platoon organization} as a solution to enhance CAV platooning without degrading the traffic performance. Under the strategy, we estimate if the traffic can absorb the disturbances from additional lane changes by measuring the local flow and speed. If flow is below and speed is above the given thresholds, the CAVs execute platoon organization. If not, the CAVs refrain from platoon organization and avoid traffic disruption. We simulate the Flow-Aware strategy of platoon organization on a realistic freeway network (State Route 99 in California) and show that CAVs under this strategy indeed form longer platoons with no disruption to the traffic flow.      
    \item The traffic capacity cannot increase further by forming longer CAV platoons on the road and reducing the headways within the platoons. Based on the queuing theory, an output flow is bounded by the input flow in a closed system. However, the input flow is composed of CAV platoons that are not longer, as platoon organization is yet to be performed. Therefore, the output flow is bounded by the input flow, i.e., the capacity cannot increase further by the maneuvers of CAVs on the road. 
\end{itemize}

The CAVs may be motivated to form longer CAV platoons regardless of the capacity increase, due to the energy efficiency improvement \cite{Muratori2017}. There is an ongoing research to validate the improvement of energy efficiency in passenger vehicles by driving in platoons. For instance, Altinisik et al experimented with a platoon of two passenger cars and found a significant reduction in air drag in the leading vehicle \cite{Altinisik2015}. Kaluva et al computed that a longer platoon reduced the average air drag coefficient of the platoon, saving energy \cite{Kaluva2020}. Also, Liu et al show that vehicles driving in connectivity can improve traffic stability and fuel efficiency \cite{Liu2020}. To facilitate the formation of long CAV platoons in a mixed traffic, we advocate for the Flow-Aware strategy of platoon organization to be implemented.

We believe our work can improve. In this study, we assume that lane changes are executed manually by human drivers in CAVs. It may be difficult for human drivers to implement the Flow-Aware strategy of platoon organization, which is based on the measurement of the local traffic condition. Because we do not know how well the human drivers will execute the platoon organization strategy, the estimated traffic performance may be biased. This bias can be overcome by assuming that the lane changes will be automated in the future and applying a model for the automated lane changing to execute the platoon organizations strategy. There is an on-going research on the lateral controller for automated vehicles. Although most works focus on safety and efficiency \cite{Wang2019}, some develop automated lane change controllers that consider the traffic flow. For instance, Wang et al \cite{Wang2016} propose a centralized controller for cooperative lane changes that explicitly models the movement of vehicle upstream in the target lane. Their controller reduces the braking and waiting times of upstream vehicles following the lane-changing vehicle. The model for automated lane changing can replace the human-driving model for lane changing and better estimate the traffic impact of platoon organization.




Also, we lack the model to capture the preference of CAV drivers in choosing between traveling in a long CAV platoon or traveling in a faster lane without a platoon. It is possible that the adjacent lane of a CAV presents a higher probability of forming longer platoons, but flows at a lower speed than the current lane. It is also possible that the adjacent lane of a CAV flows faster than the current lane, but the CAV is travelling in a platoon already. In this paper, we assume that under the PD lane strategy, the CAVs change lanes to reach the PD lane regardless of the speed difference between the current lane and the target lanes (although if the target lane is very crowded, the CAV is less likely to succeed in joining the target lane). A proper model of the CAV behavior in choosing between driving in a platoon and driving faster will better estimate the induction of lane changes under platoon organization. However, we would like to note that the Flow-Aware PD lane strategy is designed with a condition of the local speed in Equation \eqref{flow-aware-condition}. A CAV performs platoon organization only when all lanes have a very high speed (above 90km/hr in this paper). Therefore, the lack of a preference model between a longer platoon and a faster speed plays an insignificant role in the validation results of the Flow-Aware strategy.

There are many ways to improve the Flow-Aware strategy of platoon organization. First, we calibrated the threshold values, $\rho_q$ and $\rho_v$, in Equation \eqref{flow-aware-condition} by trial and error. We used only one value of flow threshold (vph/lane) for the entire network at a given CAV penetration, though the shock of lane changes may be absorbed differently at locations with various road geometries. The future research can design methods to calibrate the thresholds in a systematic manner that can be applied to various roads geometries.

Second, we assumed the availability of local measurements on vehicle count and speed and used these data to evaluate the condition in Equation \eqref{flow-aware-condition}. In practice, such data may not be available so that the CAVs need to use other sources of data to find if local traffic can absorb the shock of more lane changes. It's also possible that more sophisticated data will be available, such as individual vehicle data like gap, acceleration, and desired lane to travel. Different sources of data can be used to represent the traffic state for the Flow-Aware strategy of platoon organization with improved formulation of the condition in Equation \eqref{flow-aware-condition}.

Third, we can add sophistication to the Flow-Aware strategy, for instance by developing a learning controller that better models the complex dynamics of how a lane change can impact the traffic flow. The algorithm can model the traffic flow outcome of a lane change given various states measured by the sensors in CAVs, such as current gap, speed, acceleration, lane, and platoon length. This controller can replace the heuristic condition developed in the Equation \eqref{flow-aware-condition} and perhaps allow a longer platoon length than the Flow-Aware PD lane strategy. A useful resource is `Flow', an open-source deep learning platform developed in University of California, Berkeley, which supports microscopic traffic simulation as an environment \cite{wu2017}.

In addition, the evaluation of the energy efficiency with platoon organization is missing in this paper. It is possible that platoon organization will improve the energy efficiency of individual CAVs, while they systematically deteriorate the energy efficiency by creating congestion. If so, the Flow-Aware strategy must be implemented not only to ensure maximal traffic flow but also to avoid energy inefficiency as a whole. The future research must investigate the energy efficiency trade-off between individual vehicles and the traffic system under various operation scenarios of CAVs.


\section{Appendix} 
\label{sec:appendix}

\subsection{The Microscopic Traffic Model}
\label{appendix:micro_model}

This section briefly describes the microscopic traffic model from \cite{Liu2018_Modeling} used in this study. The model has car-following algorithms and the lane-change algorithms. The car-following algorithms are defined separately for human driving and automated driving with CACC. The algorithm for human drivers calculates the desired acceleration as the minimum of three values – acceleration from Newell’s simplified car following model \cite{Newell2002}, Intelligent Driver Model’s free flow acceleration \cite{Treiber2000}, and Gipps acceleration with safe distance \cite{Ciuffo2012}. The algorithm for automated vehicles calculates the acceleration from a model calibrated with experimental data of production vehicles instrumented with CACC \cite{Milans2014}. 

The lane-change algorithms in this model are similar for human driving and automated driving with CACC because the CACC does not provide automation for lane-changing task. In other words, human drivers in both CACC and non-CACC vehicles perform lane-change movements manually. However, the motivation for discretionary lane changes are different between the human driver and the CACC vehicles. The CACC vehicles is assumed to have a stronger motivation to stay in a platoon than to seek for higher speeds. In addition, the algorithms can emulate the complex lane-change behaviors, such as anticipatory lane changes in response to a slow speed downstream, cooperative behaviors to slow down to increase a gap for the lane-changers, and recovery responses after letting in a lane-changer with smaller reaction times and headway. 

\subsection{Supplementary Results on Flow-Aware Platoon Organization}
\label{appendix:flowaware-results}

\begin{table*}[width=\textwidth,cols=6, pos=H]
\caption{Increase of Lane Changes and Platoon Lengths without Flow Disruption under Flow-Aware Platoon Organization (Supplementary Result to Fig.~\ref{fig:totalcomparison})}
\label{tab:network_result_v3}
\centering 
\begin{tabular*}{\textwidth}{@{\extracolsep{\fill}}llllll}
\toprule   
\makecell[l]{CAV\\Penetration\\in \%} &  \makecell[l]{Network\\Demand\\Scale Factor} &\makecell[l]{Platoon\\Organization}  & \makecell[l]{Network\\Discharge Flow\\in vph$^*$} &   \makecell[l]{Average Number of\\Lane Changes\\in count/veh$^*$} & \makecell[l]{Average\\Platoon Length, $\Bar{L}$, \\in CAV count$^{*}$}   \\
\midrule
0 & 100\% &  \makecell[l]{Baseline} & 10,675.8 &   2.40 & NA    \\ 
\midrule
\multirow{3}{*}{\makecell[l]{20}}  & \multirow{3}{*}{100\%} & \makecell[l]{Baseline}  & 10,687.3 &   2.35 & 1.87      \\
\\[-1em]
\cline{3-6}
\\[-1em]
 &  & \makecell[l]{PD}  & {\color{gray}10,675.8} (-0.1\%)  & {\color{gray}3.40} (+45.0\%) & {\color{gray}4.47} (+138.9\%)   \\ 
\\[-1em]
\cline{3-6}
\\[-1em]
 &  & \makecell[l]{Flow-Aware PD}  & {\color{gray}10,687.8} (+0.0\%)  & {\color{gray}2.58} (+9.7\%) & {\color{gray}2.25} (+20.0\%)   \\ 
\midrule
\multirow{3}{*}{\makecell[l]{40}} & \multirow{3}{*}{110\%} & \makecell[l]{Baseline} &  11,739.8    & 2.40 & 3.28   \\
\\[-1em]
\cline{3-6}
\\[-1em]
 &   & \makecell[l]{PD}  & {\color{gray}11,402.0} (-2.9\%) & {\color{gray}4.90} (+103.9\%) &  {\color{gray}6.81} (+107.2\%) \\ 
\\[-1em]
\cline{3-6}
\\[-1em]
 &   & \makecell[l]{Flow-Aware PD}  & {\color{gray}11,733.8} (-0.1\%) & {\color{gray}2.93} (+22.0\%) &  {\color{gray}3.94} (+20.0\%) \\ 
\midrule
\multirow{3}{*}{\makecell[l]{60}}  & \multirow{3}{*}{117\%}  &  \makecell[l]{Baseline} & 12,449.3  &   2.15 & 4.27   \\
\\[-1em]
\cline{3-6}
\\[-1em]
 & & \makecell[l]{PD}   & {\color{gray}11,780.3} (-5.4\%)  & {\color{gray}5.55} (+157.6\%) & {\color{gray}7.12} (+66.6\%)    \\ 
\\[-1em]
\cline{3-6}
\\[-1em]
 & & \makecell[l]{Flow-Aware PD}   & {\color{gray}12,446.8} (-0.0\%)  & {\color{gray}3.06} (+42.2\%) & {\color{gray}4.98} (+16.6\%)    \\ 
\midrule
\multirow{3}{*}{\makecell[l]{80}}  &  \multirow{3}{*}{125\%} & \makecell[l]{Baseline}  & 13,347.0   & 2.03 & 5.97     \\
\\[-1em]
\cline{3-6}
\\[-1em]
& &\makecell[l]{PD}   & {\color{gray}12,378.8} (-7.3\%)  & {\color{gray}5.24} (+157.9\%) & {\color{gray}7.78} (+30.4\%)  \\ 
\\[-1em]
\cline{3-6}
\\[-1em]
& &\makecell[l]{Flow-Aware PD}   & {\color{gray}13,356.8} (+0.1\%)  & {\color{gray}2.66} (+30.9\%) & {\color{gray}6.47} (+8.5\%)  \\ 
\midrule
100 &  {127\%} & Baseline & 13,069.0   & 1.82 & 7.55\\
\bottomrule  
\multicolumn{6}{l}{\makecell[l]{$^*$In parenthesis is the percentage difference to the baseline result, i.e., (Current result - Baseline) / Baseline $\cdot$ 100\%.}}
\end{tabular*}
\end{table*} 

Table.~\ref{tab:network_result_v3} presents the detailed results visualized in Fig.~\ref{fig:totalcomparison}. PD indicates the PD lane strategy without considering the traffic state. Flow-Aware PD indicates the Flow-Aware strategy of PD lane that considers the traffic state by condition in Equation \eqref{flow-aware-condition}.

\printcredits
\section*{Acknowledgement} 
We thank Professor Carlos Daganzo, Professor Michael Cassidy, Professor David Kan, Doctor Hao Liu, and Doctor Xiao-Yun Lu at University of California, Berkeley, for their wonderful guidance and insight throughout this research.  


\bibliography{bibliography}


\bio{}
S. Woo is currently a Ph.D.  student  in  civil  and  environmental  engineering from the University of California, Berkeley. Her research focuses on the optimal operation and planning of transportation and energy systems  that  fully  utilize the benefit of connected, automated, and electrified vehicles.
\endbio

\bio{}
A. Skabardonis is currently a Professor of Civil and environmental engineering at the University of California, Berkeley. He is an internationally recognized expert in trafﬁc ﬂow theory and models, trafﬁc management and control systems, design, operation and analysis of transportation facilities, intelligent transportation systems, energy, and environmental impacts of transportation.
\endbio


\end{document}